# Interfacial Dzyaloshinskii–Moriya interaction and chiral magnetic textures in a ferrimagnetic insulator

**Shilei Ding** [1,2,3,‡], **Andrew Ross** [2,3,‡], **Romain Lebrun** [2], **Sven Becker** [2], **Kyujoon Lee** [2], **Isabella Boventer** [2,4], **Souvik Das** [2,3], **Yuichiro Kurokawa** [2,5], **Shruti Gupta** [2], **Jinbo Yang** [1,6,7], **Gerhard Jakob** [2,3], **Mathias Kläui** [2,3,8]

[1]*State Key Laboratory for Mesoscopic Physics, School of Physics, Peking University, Beijing 100871, China*
[2]*Institute of Physics, Johannes Gutenberg-University Mainz, Staudingerweg 7, 55128 Mainz, Germany*
[3]*Graduate School of Excellence Materials Science in Mainz, 55128 Mainz, Germany*
[4]*Institute of Physics, Karlsruhe Institute of Technology, 76131 Karlsruhe, Germany*
[5]*Graduate School of Information Science and Electrical Engineering, Kyushu University, Fukuoka 819-0395, Japan*
[6]*Collaborative Innovation Center of Quantum Matter, Beijing, 100871, P.R. China.*
[7]*Beijing Key Laboratory for Magnetoelectric Materials and Devices, Beijing 100871, P. R. China.*
[8]*Center for Quantum Spintronics, Department of Physics, Norwegian University of Science and Technology, NO-7491 Trondheim, Norway*

‡ These two authors contribute equally to this work

**The interfacial Dzyaloshinskii–Moriya interaction (DMI) in multilayers of heavy metal and ferrimagnetic metals enables the stabilization of novel chiral spin structures such as skyrmions. Magnetic insulators, on the other hand can exhibit enhanced dynamics and properties such as lower magnetic damping and therefore it is of interest to combine the properties enabled by interfacial DMI with insulating systems. Here, we demonstrate the presence of interfacial DMI in heterostructures that include insulating magnetic layers. We use perpendicularly magnetized insulating thulium iron garnet (TmIG) films capped by the heavy metal platinum, grown on gadolinium gallium garnet (GGG) substrates, and find a surprisingly strong interfacial DMI that, combined with spin-orbit torque results, in efficient switching. The interfacial origin is confirmed through thickness dependence measurements of the DMI, revealing the characteristic 1/thickness dependence. We combine chiral spin structures and spin-orbit torques for efficient switching and identify skyrmions that allow us to establish the GGG / TmIG interface as the possible origin of the DMI.**

The Dzyaloshinskii–Moriya interaction (DMI), an anti-symmetric exchange interaction, has been intensely studied due to the formation of chiral spin textures such as magnetic skyrmions including lattices in bulk or thin film systems [1,2] and chiral domain walls [3,4]. In bilayer systems with perpendicularly magnetized ferrimagnetic (FM) films in contact with a thin heavy metal (HM) layer, an interfacial DMI can arise based on a large spin-orbit coupling (SOC) and the lack of inversion symmetry at the interface [5]. In order to realize functional devices based on the synchronous motion of domain walls, the walls need to have a fixed chirality stabilized by interfacial DMI. This favors chiral Néel-type domain walls [3,4]. The combination of DMI and spin-orbit torques (SOT) in systems with structural inversion asymmetry leads to device concepts that are part of what is now termed spin-orbitronics [6,7].

The presence of interfacial DMI has attracted great interest recently, especially in the systems with perpendicular magnetic anisotropy (PMA), but has until recently only been identified in heavy metal and ferrimagnetic metal systems [8,9]. While bulk DMI has been observed in insulator bulk systems [10,11], interface-induced chiral spin structures have only very recently been claimed in magnetic insulators [12,13,14]. By utilizing materials exhibiting PMA, which has recently been established in various rare-earth garnets [15,16], increased packing densities of magnetic structures can be obtained, whilst the control of the magnetic state can be realized through interfacial spin-orbit torque switching at low current densities [17,18]. It has been shown that in metallic ferrimagnets, the ultra-efficient spin torque switching of DMI-induced chiral spin structures is realized [19]. However, with low magnetic damping, long spin diffusion lengths and tunable compensation points, insulating rare-earth iron-based garnets are even more promising candidates for future spin-orbitronics applications. To enable the use of ferrimagnetic insulators for spin-orbitronics, the obvious missing step is to identify and characterize possible interfacial DMI that might be realized in a heterostructure of a ferrimagnetic insulator that exhibits also PMA and efficient switching by spin-orbit torques.

In this paper, we use bilayers of the insulating garnet TmIG and platinum (Pt) to study spin-orbitronic effects in an insulating magnetic system. We first demonstrate efficient electrical switching of the magnetization by spin-orbit torques and then we identify the presence of DMI by studying the reversal of the magnetization as a function of in-plane field strength. By varying the thickness of the magnetic layer, we identify the signature of the antisymmetric DMI that scales with the inverse thickness of the TmIG layer and thus indicates an interfacial origin. Finally, we show by direct imaging that skyrmions can be present even in TmIG films without any Pt capping layer, which allows us to identify as the origin of interfacial DMI as the interface of GGG with TmIG.

The insulating rare-earth garnet, thulium iron garnet ($Tm_3Fe_5O_{12}$; TmIG) was grown on <111>-oriented gadolinium gallium garnet ($Gd_3Ga_5O_{12}$, GGG) substrates by pulsed laser deposition and structured with platinum Hall bars (GGG / TmIG / Pt, see methods for details). In Figure 1 (a), we present an exemplary X-ray diffraction (XRD) curve of a 25 nm film around the GGG (444) peak, which demonstrates a fully strained film grown on GGG, with Laue fringes indicating the high quality of sample growth. Polar optical Kerr (MOKE) microscope measurements of films consisting of 25 nm and 2.8 nm TmIG are shown in Figure 1 (b), both showing a square hysteresis loop with a strong out-of-plane remanence, confirming the PMA in our samples across the thickness range investigated. The coercivity is around 1 mT, and the saturation magnetization, $M_s$, is measured by SQUID to be 110±5 emu/cm$^3$ for the 25 nm sample which is close to the bulk value [20], while

$M_s$ for the 2.8 nm sample is found to be 50±25 emu/cm³ likely due to the reduced thickness. The $M_s$ of TmIG sample decreases as the thickness gets thinner [21], which is consistent with previous reports [22]. In all samples, we observe a clear domain structure using MOKE measurements showing that stable multi-domain states are possible in the range of TmIG thicknesses studied.

The electrical manipulation of the magnetization in such garnet heterostructures can be obtained via SOTs generated in an adjacent HM layer with a large spin-orbit coupling [16,18]. Additionally, precise knowledge of the magnetic state of magnetic insulators can be obtained by measuring the anomalous Hall effect (AHE) in a transverse voltage configuration. From this voltage, we are able to extract the magnetic state also through an effect known as spin Hall magnetoresistance [23]. The Pt Hall bar geometry used to probe the magnetic state of the TmIG layer is depicted in Figure 2 (c). We apply a probing charge current, $j = 9.5 \times 10^8$ Am$^{-2}$ and record the transverse Hall resistance as a function of the external magnetic field applied perpendicular to the sample plane $H_z$ in Figure 2 (a). The observed Hall resistance shows a well-defined, square hysteresis loop with a coercivity of $2.8 \pm 0.1$ mT. We attribute the increase in coercivity after patterning the continuous film to geometrical confinement effects [18]. We reverse the polarity of the current to exclude any polarity dependent bias effects and find the resistance change between the two magnetic states to be $\Delta R^{SMR,AHE} = 0.40 \pm 0.02$ mΩ. The ordinary Hall effect (OHE) gives a linear background, which is subtracted in Figure 2 (a), and the value was found to be $R^{OHE} = 3.5 \pm 0.2$ mΩT$^{-1}$ [24].

Next, we demonstrate efficient deterministic electrical switching using electrical pulses. In a bilayer of GGG / TmIG (2.8 nm) / Pt (7 nm), we observe a threshold current density of $j = 2.86 \times 10^{10}$ Am$^{-2}$ when applying 5 ms current pulses with an in-plane field of 2.5 mT field applied parallel to the current direction. The polarity of the pulse train is repeatedly reversed, with the transverse Hall resistance measured both before and after. We observe a variation of $R_H$ that is comparable with the variation of the transverse resistance seen during magnetization reversals by field. This indicates the complete reversal of the magnetization induced by the current. The deterministic switching is governed by the current direction and inverts with reversal of the in-plane field, which is the typical characteristic behavior of SOT switching in PMA films [25,26]. Figure 2 (d) summarizes the magnetic switching phase diagram depicting the minimum current density required for reliable, deterministic switching of the magnetic state for a given in-plane field. The change of switching field as a function of injected current density demonstrates the equivalency of the field and current, where the efficiency is defined as [27]:

$$\chi = \left|\frac{\mu_0 \Delta H}{\Delta J}\right| \qquad (1)$$

Assuming a uniform current distribution, we obtain $\chi_{j=4\times10^{10} \text{ Am}^{-2}} = (0.7 - 1) \times 10^{-12}$ mT/Am$^{-2}$, which is comparable to the efficient switching observed in Co/Pt systems [28].

Having established deterministic SOT-induced switching of our films, we next move to the central task of exploring interfacial chiral coupling effects. Such effects can be determined from an analysis of the hysteresis loop of switching a Hall cross with combined in-plane and out-of-plane fields [29]. This is shown exemplarily in Figure 3 (a) as a horizontal displacement of the hysteresis loop caused by an effective interfacial field that can be probed by varying external in-plane field, $H_x$, in combination with varying the current density in the HM layer. The probing current in the Pt generates a damping-like spin-orbit torque at the TmIG / Pt interface proportional to $\mathbf{m} \times (\mathbf{m} \times \boldsymbol{\sigma})$, where $\boldsymbol{\sigma}$ is the polarization of the spin current generated by the spin Hall effect and $\mathbf{m}$ is the magnetization vector in the TmIG film. This damping-like torque presents itself as an additional out-of-plane field inside the Néel type domain walls [8], which favors one type of domain over the other. This thus leads to a horizontal shift of the hysteresis loop as a function of the out-of-plane field, the magnitude of which increases with the in-plane field before saturating at some point. Using this method, we can thus determine both the DMI and the spin-orbit torque efficiency by measuring the shift of the hysteresis loop as a function of the current density for differing values of $H_x$. Figure 3 (a) shows such a shifted hysteresis loop in the presence of a charge current of $8.6 \times 10^{10}$ Am$^{-2}$ for the GGG / TmIG (2.8 nm) / Pt (7 nm) sample, where reversing the polarity of the probing current leads to a negative displacement of the hysteresis. The efficiency of the spin-orbit torques, $\chi$, acting on the magnetization of the TmIG is then extracted from the relation between the horizontal shift of the hysteresis as a function of probing current density, $j$.

In a system without interfacial DMI, no SOT effects will be seen as this system favours the presence of Bloch-type domain walls. Only if an in-plane field is applied along **x**, will a Néel component emerge parallel to the field, which would lead to an extractable spin-orbit torque efficiency. However, in the presence of interfacial DMI, the system stabilizes only Néel domain walls with a preferred chirality (determined by the sign of the DMI) [4]. This DMI field thus lowers the switching SOT efficiency but by applying an external field, we can compensate for the DMI field, enhancing the spin-orbit torque efficiency of the switching mechanism [29,30]. When the applied field by far exceeds the effective DMI field $H_{DMI}$, the magnetization inside the domain walls aligns with the external field. In such a case, domain walls between up-down and down-up domains have opposite chirality, and thus domain expansion is favored over domain wall motion, enhancing the efficiency of the switching [29,31]. The slope of $\chi$, Figure 3 (b), is linear for small $H_x$ before saturating at larger fields, from which we estimate a saturation field $H_{DMI} = 20 \pm 1$ mT and the maximum SOT efficiency of $\chi_{eff} = 8.1$ mT/$10^{12}$Am$^{-2}$ which can be compared to the value extracted from magnetic switching phase diagram with Eq. (1). We can estimate the effective DMI exchange constant $D$ from the saturation field $H_{DMI}$ according to [8,31]:

$$|D| = \mu_0 M_s \Delta |H_{DMI}| \qquad (2)$$

$\mu_0$ is the vacuum permeability, $M_s$ is the saturation magnetization and $\Delta$ is the width of the domain wall, which can be calculated from $\Delta = (A/K_u)^{1/2}$ [32]. Here, $A$ is the exchange stiffness where the value is estimated to be 2.3 pJ/m [33], and $K_u$ is the anisotropy energy of TmIG which can be obtained from effective anisotropy field and saturation magnetization via $H_k = \frac{2|K_u|}{\mu_0 M_s}$. The effective DMI constant obtained for 2.8 nm of TmIG is found to be $|D| = 3.6 \pm 2.0 \times 10^{-2}$ mJ/m$^2$. Considering the threshold DMI constant to stabilize Néel domain wall scales with $M_s^2$: $|D_{th}| = 2tIn(2)\mu_0 M_s^2/\pi^2$ ($t$ is the thickness of the magnetic layer) [13], the value of $|D_{th}|$ for 2.8 nm TmIG sample is calculated to be $1.2 \pm 0.6 \times 10^{-3}$ mJ/m$^2$. The value is two orders of magnitude lower than that of metal systems due to the small $M_s$ of the garnets. Since our determined effective DMI constant is larger than the threshold DMI constant, we conclude that Néel walls are favored in our TmIG system. To understand the origin of the DMI field and in particular ascertain whether it results from an interface, the key measurement is to probe the DMI as a function of thickness. A series of samples with thicknesses ranging from 2.8 nm to 16.6 nm were grown, measured and the effective DMI constant was extracted. We find that the value of the DMI constant scales linearly with the inverse of the thickness, as shown in Figure 4 demonstrating its interfacial origin [34]. The identification of an interfacial DMI is unprecedented in an insulating magnetic system, where no conduction electrons exist to mediate the exchange between the magnetic ions. To understand the origin, we first note that bulk DMI has not been reported in TmIG, and is not expected as the cubic structure of TmIG exhibits inversion symmetry. However, such inversion symmetry can be broken at the either TmIG / Pt or GGG / TmIG interface. To independently check if the DMI induces chiral spin structures, we next probe current driven domain wall motion with kerr microscope as exemplified in Figure 5(a) for a GGG / TmIG (8 nm) / Pt (7 nm) sample. The domain walls in the TmIG move to right and left according to the direction of current pulse, which implies the existence of chiral domain walls that are stabilized by DMI. By applying an in-plane field that overcomes the DMI field, the motion direction can be further changed. Finally for a sufficiently large DMI, we expect that skyrmion spin structures can be stabilized. In the simplest approximation, this holds for a critical DMI constant $D_c = 4\sqrt{AK_u}/\pi$ [35], and the skyrmion will be stabilized if $D > D_c$. Taking into account Eq. (2), we obtain the critical condition to be $\frac{H_{DMI}}{H_k} > \frac{2}{\pi}$ when skyrmion formation becomes energetically favourable. Based on the measured values of the magnetic parameters [36], we expect that for thin TmIG thus skyrmions should be stable. Indeed we observe such circular spin structures with a diameter of ~2 μm alongside worm domains after the application of an in-plane field in 8 nm of TmIG, see Fig. 5(b). Whilst such magnetic structures may be stabilized by dipolar interactions, the observation of DMI through both electrical transport measurements and chiral domain wall motion in the same sample indicate that these circular domains are also chiral in nature thus exhibiting a non-trivial Skyrmion topology. Given the fact that the skyrmion-like structures are observed even in samples without a Pt capping layer, we could finally conclude that significant interfacial DMI in the heterostructure should originates from the GGG/TmIG interface [13].

In conclusion, we have demonstrated the two key spin-orbitronic effects in a heterostructure of Pt and TmIG: We realize efficient spin-orbit torque switching with low current densities due to interfacial damping like spin-orbit torques. And in particular, we establish the presence of chiral antisymmetric exchange coupling in this GGG / TmIG / Pt system. The observed inverse proportional scaling between TmIG thicknesses and DMI strength demonstrates the interfacial origin of the DMI. The DMI stabilizes skyrmions, which are observed in GGG / TmIG bilayers highlighting that the interface between the TmIG and the GGG should generates sizeable DMI. Our results demonstrate that the necessary ingredients (SOTs, PMA, DMI and magnetic skyrmions) for new spin-orbitronic devices exist in rare-earth garnet heterostructures, thus paving the way for novel applications using this exciting class of insulating materials.

**Methods**

Sample fabrication: TmIG films were deposited on 5 mm x 5 mm (111)-oriented gadolinium gallium garnet ($Gd_3Ga_5O_{12}$, GGG) substrate by pulsed laser deposition (PLD) at a laser energy of ~1 J cm$^{-2}$ with a laser wave length of 248 nm. The distance between target and substrate is 5.5 cm, and the deposition temperature is set to be 800 °C with the oxygen pressure of 0.27 mbar. No annealing process was used and the cooling occurred at -25°C/min under 0.27 mbar oxygen pressure. 7 nm Pt was sputtered ex-situ with the sputter rate of 0.08 nm/s in an argon atmosphere. The Hall devices were patterned by electron beam lithography and subsequent Ar ion-milling. The width of the hall bar is designed to be 75 um for SOT-switching and the width of the hall bar was structured to 1 um for DMI and SOT efficiency measurement.

Characterization: The magnetic properties of TmIG originate from super-exchange coupling of the iron sub-lattices with the magnetic easy axis lying along <111> arising from negative magnetostriction and negative anisotropy constant, $K_1$ [37]. In thin films, we obtain a PMA nature of the films originating from an in-plane strain between the film and substrate. The crystal structure and strain conditions were confirmed by X-ray diffraction (XRD) and reciprocal space mapping. Reciprocal space mapping of 25 nm TmIG confirms the result of in-plane coherent strain [38], and the thinner samples are expected to be strained as well.

Magnetic and electric measurements: In-plane and out-of-plane hysteresis loops were performed by Kerr microscopy and SQUID. The domain structure was measured by Kerr microscopy. All of the electric measurements were performed at room temperature, probe and pulse currents were provided by a Keithley 2400, and a Keithley 2182A nanovoltmeter was used to measure the transverse Hall voltage. In the switching phase diagram measurement, the maximum current was set to be 24 mA (the current density is $4.39 \times 10^{10}$ Am$^{-2}$) to avoid damaging the device.

During the preparation and submission of this manuscript, we became aware of other studies by Vélez et al. [12] and C. O. Avci et al. [13] showing results of domain wall motion in TmIG/Pt and Q. M. Shao et al. reporting a skyrmion lattice inferred from indirect topological Hall effect measurements at elevated temperatures in TmIG/Pt [14], which are in line with our direct observation of skyrmions.

**Acknowledgments**

We acknowledge support by the Graduate School of Excellence, Material Science in Mainz (GSC 266); the German Research Foundation DFG (Project numbers 290396061/TRR173, 358671374, 403502522); the JSPS Program for Fostering Globally Talented Researchers; the National Key Research and Development Program of China (2016YFB0700901, 2017YFA0206303, 2017YFA0403701) and the National Natural Science Foundation of China (51731001,11975035, 11805006) R.L. acknowledges the European Union's Horizon 2020 research and innovation programme under the Marie Skłodowska-Curie grant agreement FAST number 752195

**Figure captions**

Figure 1. Structure and magnetic properties of TmIG. (a) X-ray diffraction measurements of 25 nm thick TmIG layers around the 444 peak of GGG. (b) The out of plane hysteresis loops of 25 nm and 2.8 nm thick TmIG samples, obtained from the magnetic contrast using a Kerr microscope.

Figure 2. The deterministic current induced switching of a 2.8 nm thick TmIG structure. (a) Anomalous Hall resistance measurement of 2.8 nm TmIG / Pt as a function of out of plane field $H_z$. The Hall resistance is a direct measure of the relative orientation of the magnetization vector of the TmIG, resulting in a hysteresis with a coercivity of $2.8 \pm 0.1$ mT. (b) Current induced magnetic switching for an in-plane field $H_x = 2.5$ mT with trains of 3, 5 ms-long current pulses of $j = 2.86 \times 10^{10}$ Am$^{-2}$. The switching orientation changes sign with an external field or current pulse polarity. (c) Hall device schematic and the relevant coordinates. (d) Magnetic switching diagram measured by applying current pulses for varying applied $H_x$ from 0 to 3 mT showing the minimum current needed for complete, deterministic switching.

Figure 3. Measurement of the effective DMI constant and SOT efficiency in GGG / TmIG (8 nm) / Pt (7 nm). (a) AHE measurement for a probe current of $8.6 \times 10^{10}$ Am$^{-2}$ with a bias in-plane field $H_x$ of 14 mT. The horizontal shift of the hysteresis loops corresponds to the SOT induced effective field. (b) SOT efficiency calculated from the horizontal displacement for different probing current densities at different in-plane fields $H_x$. $\chi_{eff}$ is the maximum SOT efficiency, and $H_{DMI}$ is the field at which $\chi$ saturates.

Figure 4. The effective DMI constant as a function of $1/t_{TmIG}$. The linear relation reveals the interfacial origin of the measured DMI.

Figure 5. (a) Example of the current driven domain wall motion in GGG / TmIG (8 nm) / Pt (7 nm). (b) Skyrmions distributed in a 8 nm thick TmIG sample obtained at 0 mT out-of-plane filed after in-plane field saturation (the circles with strong black-white contrast are structural defects where artificial contrast is generated due to the imaging procedure).